\documentclass[preprint,12pt]{elsarticle}
\usepackage{amssymb}
\usepackage[]{epsfig}
\usepackage{amsmath,graphicx}
\usepackage{pstricks}
\journal{Physics Letters B}
\begin{document}
\begin{frontmatter}

\title{ Horizons and  the  Wave Function of Planckian Quantum black holes}
\author{Euro Spallucci\footnote{e-mail:Euro.Spallucci@ts.infn.it} $\,$\footnote{Senior Associate}}
\address{INFN,\\
Sezione di Trieste, Trieste, Italy}
\author{Anais Smailagic\footnote{e-mail:Anais.Smailagic@ts.infn.it} $\,$\footnote{Senior Associate}}
\address{INFN, Sezione di Trieste, Italy}

\begin{abstract}
At the Planck scale the distinction between elementary particles and black holes becomes fuzzy. 
The very definition of a "quantum black hole" (QBH) is an open issue. Starting from the idea
that, at the Planck scale, the radius of the event horizon undergoes quantum oscillations, we 
introduce a black hole mass-radius Generalized Uncertainty Principle (GUP) and derive
a corresponding \emph{gravitational wavelength}. 
Next we recover a  GUP encoding effective geometry. This semi-classical gravitational description admits black hole configurations
only for masses higher than the Planck mass. Quantum corrections lead to a vanishing  Hawking temperature 
when the Planck mass is approached from above. Finally we replace our semi-classical model by 
 a relativistic wave equation for the  \emph{ "~horizon wave function~"}. The solution admits a discrete mass spectrum which is bounded from below by a stable ground state with energy close to the Planck mass. Interestingly higher angular momentum states fit onto  Regge trajectories indicating their stringy nature.
\end{abstract}
\end{frontmatter}

\section{Introduction}
\label{intro}
The "point-particle" is the simplest way to represent   microscopic objects. It corresponds to a mathematical point endowed with physical properties like mass, charge, etc. Such an idealized representation of  elementary particles is to be taken
with  due caution. True physical objects cannot have zero volume nor infinite density. The experimental side-  the term 
point-particle is justified for objects  whose linear size is  below the resolution power of
the instrument used to probe them.\\
The very existence of fundamental (dimensional) constants like  $c$-speed of light, $G_N$-Newton constant, $\hbar=h/2\pi$ -(reduced) Planck constant, enables one to combine them into two fundamental length scales. By conveniently setting $c=1$ one defines

\begin{itemize}
\item $r_g\equiv m\,G_N$, which is the gravitational length scale . 
In this range of length   gravitation is "strong"  and closed trapping surfaces, colloquially referred to as  "horizons" \footnote{It is common practice to define $r_G$ as $r_G=2mG_N$,  is the Schwarzschild radius of a "classical" black hole. We shall show  that, near the Planck scale, the QBH radius is actually one half of the
Schwarzschild radius.} , can appear. Whenever a physical  falls trough a "hoop" of radius 
$r_s$ it will unavoidably collapse under its own weight  into a black hole. This is the so-called \emph{hoop conjecture} introduced by K.Thorne \cite{thorne}.
\item $\lambda_C \equiv\hbar/m$  is the Compton wavelength. This is the length scale where quantum effects are  dominant and a particle can only be described  in terms of its wave function.
\end{itemize}

The above length scales identify  two distinct mass regions:
\begin{itemize}
\item Black hole sector: $r_g > \lambda_C$, with $m>m_P$
\item Particle sector: $\lambda_C > r_g$, with $m<m_P$
\end{itemize}
The boundary between the two regimes is determined by the Planck mass $m_P$ and Planck length $l_P$:
\begin{eqnarray}
&& m_P^2 = \frac{\hbar}{G_N}=\frac{l^2_P}{G^2_N}\ ,\\
\end{eqnarray}
The \emph{Planck scale} is the  regime where gravity merges with quantum mechanics. An  object of mass $m<m_P$ is an ordinary quantum particle obeying known quantum mechanical rules. When its energy is above $m_P$ a quantum particle develops   
an event  horizon shielding it from an asymptotic observer and becomes quantum black hole QBH. The behavior and properties of these kind of microscopic objects deserve a thorough investigation which is still underway.  \\
This paper is organized as follows. In Section({\ref{wavel}) we introduce a \emph{new} Mass-Horizon Uncertainty Principle
and recover the quantum wavelength $\lambda_G$ for a Planckian object. In Section (\ref{g1}) we identify the two roots
of $\lambda_G$ with the radii of the inner (Cauchy) and outer (Killing) horizon of an effective geometry. A black hole configuration
is admitted only if the mass is heavier than the Planck mass. We study the thermodynamical properties of this object
and find that the Hawking temperature and the entropy both vanish at the Planck mass. This behavior is a signal of
a breakdown of the effective geometric description  and a full quantum approach is needed.
In Section(\ref{qbh}) we solve the horizon wave equation  and recover the discrete mass spectrum of a QBH.
 Section(\ref{closing}) is devoted to a brief summary of the main results obtained.

\section{The wavelength of QBH} 
\label{wavel}
At the Planck scale the distinction between particle  and black hole becomes murky.  
In quantum mechanics the position and momentum of  a  particle are conjugate variables which cannot be simultaneously measured with arbitrary
precision. Accordingly, one expects a similar behavior for a QBH. The effects of gravity 
lead to  the Generalized Uncertainty Principle (GUP) whose purpose is to put  quantum particles and QBHs on the same footing 
\cite{Maggiore:1993rv,Bambi:2007ty,Banerjee:2010sd,Carr:2015nqa,Lake:2015pma,Carr:2014mya,Casadio:2017sze,Lake:2016enn,Lake:2018hyv,Carr:2020hiz,Tawfik:2014zca,Lake:2020rwc}. \\
The GUP can be written in a most general form as \cite{Kempf:1994su}

\begin{equation}
\Delta x \Delta p \ge \frac{\hbar}{2} + \alpha G_N \left(\,\Delta p\,\right)^2+\beta\frac{\left(\,\Delta x\,\right)^2}{ G_N }\ , \label{gup}
\end{equation}
where $\alpha,\beta$ are numerical constants.
\\
A  heuristic derivation of the GUP consists in including gravity in the "Heisenberg microscope" ideal experiment where the particle position is determined by observing a scattered photon and applying simple optical principles 
\cite{Maggiore:1993rv,Adler:1999bu} . While for a particle this is a legitimate approach,
for a QBH the situation is more subtle. A photon reaching the QBH center cannot be scattered back into the microscope. Therefore the  QBH center is not directly observable being shielded by the horizon.
 In principle  one can by-pass the problem by reconstructing the location of the QBH center through the angular distribution of outgoing thermal photons during Hawking  radiation. We will show  in Section(\ref{g1}) that a QBH has a vanishing temperature as
 its mass approaches $m_P$ and the argument advocating the Hawking radiation fails.\\
  Another derivation of GUP \cite{Scardigli:2007bw} considers the  
the disturbance induced in the metric by the measuring procedure  to obtain 
\begin{equation}
\Delta x \ge \frac{\hbar}{2\Delta E} + 2G_N\Delta E
\end{equation}
In all formulations $\Delta x$  refers to the uncertainty in the \emph{ position} of the QBH center, exactly as in   the case of an ordinary point-particle. The existence of the horizon is not explicitly encoded into this form of GUP. As a matter of fact, 
when considering the evolution of a  BH one is not referring to its motion in space, but to the mass and radius variations.
However, at  the Planck scale the  horizon cannot be described  as a static, geometric boundary. It is subject to quantum fluctuations making its position uncertain.
A preliminary analysis of both classical and quantum dynamics of an oscillating horizon can be 
found in \cite{Spallucci:2016qrv}.\\
On the quantum mechanical side, let us recall the fundamental lesson from the hydrogen atom  example: the dynamics of that system is determined by the relative  motion of the electron with respect to the common center of mass, and not by the (trivial) motion of the center of mass itself. In the same way the dynamics of a QBH is uniquely referred to by its horizon fluctuations and not by its global
motion through space. 
Thus one needs to start from a version of GUP explicitly  taking into account these  fluctuations rather than the uncertainty in the position of the QBH center.\\
Furthermore the GUP has  to encode the essential difference between a particle and QBH as discussed in Section(\ref{intro}) and described by the condition  $m\ge m_P$.\\
For all these  reasons,   we propose a  version of (\ref{gup}) entitled  Horizon Uncertainty Principle (HUP), where $\Delta x$ is replaced by $\Delta r_H$ and $\Delta p$ by $\Delta m$, with $\alpha =0$ and  $\beta$ 
 a free parameter. The result is:

\begin{equation}
\Delta r_H\, \Delta m\ge \frac{\hbar}{2} +\beta\, \frac{\left(\, \Delta r_H\,\right)^2}{G_N}
\label{bhup}
\end{equation}

Relation (\ref{bhup}) imposes the lower bound condition:

\begin{equation}
\Delta m\ge \sqrt{ \frac{\hbar\beta}{G_N}  } =\sqrt{\beta}\,m_P
\end{equation}
which is in agreement with the appearance of the horizon  {\bf only} above the Planck mass. \\
In our case, the so-called Compton-Schwarzschild  wavelength $\lambda_G$ introduced in \cite{Lake:2015pma} takes the  form  

\begin{equation}
\lambda_G =\frac{\hbar}{2m} +\beta\, \frac{\lambda_G^2}{G_Nm} \label{gwave}
\end{equation}
As we are dealing with gravity, it is useful to express our results in terms  of $l_P$ and $m_P$ as fundamental constants using relations

\begin{eqnarray}
G_N= \frac{l_P}{m_P}\ ,\\
\hbar=m^2_P G_N=m_P l_P
\end{eqnarray}

 \begin{figure}[h!]
\includegraphics[width=10cm]{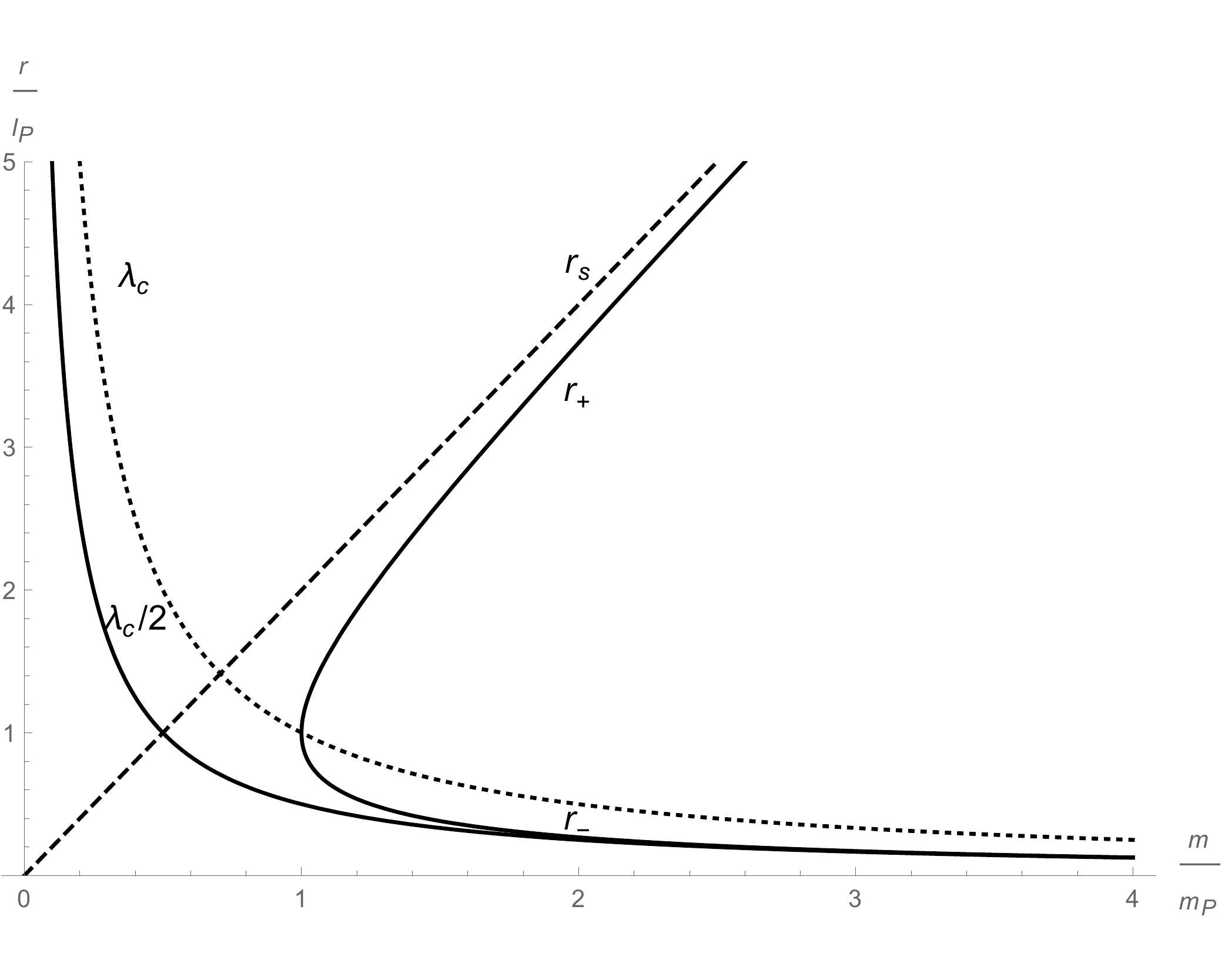}
\caption{Plot of the Compton wavelength and the two horizons by varying  $m$. The three curves meet at the
\emph{bifurcation point} $m=m_P$. Each of them approaches a different asymptote. Note that $r_+ $ differs from the classical Schwarzschild radius $r_s$  as $m\to m_P$. We remark that the region of lengths smaller than $l_P$ has no physical  meaning and only the branch above $1$  of the plotted functions is  relevant.}
\label{bif}
\end{figure}

Equation (\ref{gwave}) is a quadratic relation between $m$ and  $\lambda_G$.
Solving the equation we find

\begin{equation}
\frac{\lambda_G^\pm}{l_P} = \frac{m}{2\beta\,m_P}\left(\, 1 \pm \sqrt{1-2\beta \frac{m_P^2}{m^2}}\,\right)
\end{equation}
Again, we notice that $\lambda_G^\pm$ are real  only for $m\ge\sqrt{ 2\beta} m_P$ as required.
 If  we assume that the transition from particles to QBHs to occurs precisely at the Planck scale, then  $\beta=1/2$.\\
For $m=  m_P$ the two roots coincide $\lambda_G^+=\lambda_G^-=  l_P$. 
It is interesting to consider the \emph{large mass} limit of $\lambda_G^\pm$. Far above the Planck mass
we find
\begin{eqnarray}
&& m>> m_P\ ,\\
&& \lambda_G^+ \sim 2\,G_N\,m=r_s\ ,\\
&& \lambda_G^-= \frac{\lambda_C}{2}\ ,\\
&&\lambda_C=\frac{\,m_P}{m}\,l_P
\end{eqnarray}
 
 The resulting picture, presented graphically in Figure (\ref{bif}),  shows that by increasing the mass of a particle its Compton wavelength approaches the \emph{bifurcation point} $m=m_P$ where $\lambda_C$ splits into  $\lambda_G^+$ and $\lambda_G^-$. With further increase of $m$,   $\lambda_G^+$ approaches the classical Schartzschild radius  while  $\lambda_G^-$  decreases towards zero and merges with 
$\lambda_C/2$ in the unphysical region below the Planck length. The presence of a bifurcation point indicates
the  transition between particles and black holes. Also notice that 
$\lambda_G^-$ for any $m>m_P$ is below the Planck length and has no physical
meaning.  As a matter of fact, only  $\lambda_G^+$ is physically relevant.

\section{Effective geometry}
\label{g1}
In the previous section we introduced a QBH wavelength which smoothly interpolates between the
Compton wavelength and the Schartzschild radius. On the other hand, people are more familiar with the geometric description of BHs in terms of the space-time metric. A simple way to introduce an \emph{effective geometry} is to identify $\lambda^\pm_G$ with
the radii of the inner and outer  horizons:
\begin{eqnarray}
&& \lambda^+_G\equiv r_+= mG_N\left(\, 1 +\sqrt{1-\frac{m^2_P}{m^2}}\,\right)\ ,\label{h1}\\
&& \lambda^-_G\equiv r_-= mG_N\left(\, 1 -\sqrt{1-\frac{m^2_P}{m^2}}\,\right)\label{h2}
\end{eqnarray}

To obtain the corresponding  metric, we  rearrange HUP (\ref{bhup}) in the form 
 
  \begin{equation}
1-\frac{2G_N \Delta m}{\Delta r_H} +\frac{G_N m^2_P}{\left(\, \Delta r_H\,\right)^2}=0
\end{equation}

which is nothing but a horizon equation in geometric description. It is now easy to reproduce the corresponding line element

\begin{eqnarray}
&&ds^2=-f\left(\, r\,\right)\, dt^2-f^{-1}\left(\, r\,\right)\,dr^2 +r^2 \, d\Omega^2\label{qbh1}\\
&&f\left(\, r\,\right)=1 -\frac{2m\,G_N}{r}+\frac{\hbar G_N}{r^2} \nonumber\\
&&\hbar=G_N\,m^2_P\nonumber
\end{eqnarray}

The result is a Schwarzschild-like geometry with a quantum gravity correction $\hbar G_N /r^2$. Contrary to the standard  
 geometry, the line element (\ref{qbh1}) describes a black hole \emph{only}  for
$m\ge m_P$ as it is required for a QBH. A closer look at (\ref{qbh1}) shows that it has the same form as Reissner-Nordstr\"om BH geometry, but with the fixed value of the ``charge''  $Q^2=\hbar G_N= G_N^2\,m^2_P$.\\
It is interesting to see the impact of  quantum corrections  on the BH thermodynamics. \\
In geometric units, $c=G_N=\hbar=k_B=1$  ( $k_B$ is the Boltzman constant ), the  Hawking temperature 

\begin{equation}
T_H= \frac{1}{4\pi}\left(\,\frac{df}{dr}\,\right)_{r=r_+}=\frac{1}{4\pi r_+}\left(\, 1-\frac{l_P^2}{r_+^2}\,\right)
\end{equation}

\begin{equation}
T_H\left(\, r_+=\sqrt{3}l_P\,\right)\equiv T_{max}=\frac{1}{6\pi\sqrt{3}}\frac{1}{l_P}
\end{equation}

\begin{equation}
T_H\left(\, r_+=r_-=l_P\,\right)=0
\end{equation}

Initially, the temperature increases   towards the maximum $T_{max}$ and then quickly drops to zero for $m=m_P$. Contrary to the
general expectation, $T_H$ does not diverge at the Planck scale. Rather the black hole freezes to the extreme
degenerate configuration:  
\begin{eqnarray}
&& r_{ext}=G_N m_P=l_P\ ,\\
&& m_0= m_P  \ ,\\
&& T_H=0
\end{eqnarray}

In accordance  with the Third Law of Thermodynamics one expects the entropy to vanish for $T_H=0$.
In fact by integrating the First  Law of Thermodynamics one obtains

\begin{equation}
dm= T_H\, dS_H\longrightarrow S_H=\frac{2\pi}{G_N}\int_{r_{ext}}^{r_+} dx\, x =\frac{\pi}{G_N}\,\left(\, r^2_+-r^2_{ext}\,\right)
\end{equation}
For large BHs with $r_+>> l_P$ one recovers the standard form of the Area Law

\begin{equation}
S_H \longrightarrow \frac{\pi}{G_N} r^2_+ = \frac{A_H}{4\,G_N}
\end{equation}

while, for $r_+\to l_P$ one  obtains {\bf zero entropy}. 

Important feature of any quantum theory of gravity is the Holographic Principle (HP)
\cite{tHooft:1993dmi,Susskind:1994vu} which implies that the BH dynamics is confined solely on the horizon and not in the bulk.
It follows that BH entropy is stored on its surface,  with one bit of information per Planck cell.  Zero entropy is achieved once the horizon surface shrinks to a single cell at zero temperature.\\
The results  obtained in this section refer to a neutral, non-rotating BH, but their validity is general. In fact, in case of rotating and charged BHs, the Hawking
radiation depletes both charge and angular momentum leading finally to a Schwarzschild-like QBH.

\section{The horizon wave function.}
\label{qbh}

Classical models of evolving QBHs were introduced many years  ago in the framework of the theory of relativistic, self-gravitating
 membranes \cite{Aurilia:1989sb,Aurilia:1990bb,Aurilia:1990uq,Maggiore:1994ww,Ansoldi:1997cz}. 
Recently, particle-like models of QBHs have been considered 
\cite{Dvali:2011aa,Dvali:2012gb,Dvali:2013vxa,Dvali:2015rea}. 
In accordance with the latter, we provided  \cite{Spallucci:2016qrv} a  model  where the horizon dynamics is translated into the motion of  a relativistic point-particle trapped in self-consistent gravitational
potential . In this framework the horizon oscillations are effectively described by the periodic motion of its particle-like 
analogue. The dynamics of this representative "~particle~" is encoded in a "~relativistic Hamiltonian~" given by

\begin{equation}
\mathcal{H}\equiv \vec{p}{}^{\, 2}_H +m_H^2
\end{equation}
where $\vec{p}_H$ is the momentum, and the particle mass $m_H$ is  expressed in terms 
of the  horizon radius $r_+$. For the metric (\ref{qbh}) this function is determined by the horizon condition $f\left(\, r_H\,\right)=0$

\begin{equation}
m_H(r_+)=\frac{r_+}{2G_N}\left(\, 1 +\frac{G_N^2 m^2_P}{r_+^2}\,\right)
\end{equation}

In order to keep the notation as simple as possible, we shell drop the suffix "${}_+"$ from all the relevant
quantities keeping in mind that all of them are referring to the outer BH horizon. 
Thus, the horizon wave equation \cite{Spallucci:2016qrv} 
is\footnote{In this section we essentially deal with quantum mechanics, thus it is convenient to
use natural units: $\hbar=1\ ,c=1$, $G_N=l^2_p=m^{-2}_P$ }

\begin{equation}
\mathcal{H}\Psi\left(\,\vec{r}\,\right)=E^2\, \Psi\left(\,\vec{r}\,\right)
\end{equation}
The $O(3)$ symmetry of the problem allows to express the angular dependence
of the wave function in terms of spherical harmonics $Y_l^m \left( \, \theta\ , \phi\,\right)$. The
remaining part of the wave function  $\psi\left(\, r\,\right)$ satisfies the radial equation

\begin{equation}
\left[\, \frac{1}{r^2}\frac{d}{dr}r^2 \frac{d\psi}{dr}\right] +\left[\, E^2 -\frac{r^2}{4G_N^2}\left(\, 1 +\frac{G_N^2 m^2_P}{r^2}\,\right)^2 -\frac{l\left(\,l+1\,\right)}{r^2}\,\right]\psi\left(\,r\,\right)=0 \label{weq}
\end{equation}

 (\ref{weq}) looks like a relativistic wave equation for a "particle" moving in some effective
potential   $ V_{eff}\left(\, r\,\right)$    given by

\begin{equation}
V_{eff}\left(\, r\,\right)\equiv \frac{r^2}{4G_N} + \frac{\hbar^2}{r^2} +\frac{l\left(\, l+1\,\right)}{r^2}
\label{veff1}
\end{equation}

The first term in (\ref{veff1}) is an harmonic potential describing the vibrations of the horizon \footnote{A slightly different
approach to the horizon wave function has been proposed in 
\cite{Casadio:2013aua,Casadio:2015sda,Casadio:2016dzy} }.
The second term is a GUP contribution $\hbar=G_N\,m^2_P$ to the potential, while the third term is the usual centrifugal barrier.\\
The wave equation admits an exact solution:

\begin{equation}
\psi\left(\,r\,\right)=N_n \left(\, \frac{r^2}{2G_N}\,\right)^s e^{-r^2/4G_N} L_n^{2s +1/2}\left(\, \frac{r^2}{2G_N}\,\right)
\end{equation}
where, $L_n^{2s +1/2}\left(\, \frac{r^2}{2G_N}\,\right)$ is a generalized Laguerre polynomial; $N_n$ is te normalization
coefficient, and

\begin{equation}
s=\frac{1}{4}\left[\, \sqrt{ G_N^2 m^4_P  +\left(\, 2l+1\,\right)^2 } -1\,\right]=\frac{1}{4}\left[\, \sqrt{1  +\left(\, 2l+1\,\right)^2 } -1\,\right]
\end{equation}

As the form of $V_{eff}\left(\, r\,\right)$ suggests, one finds a discrete energy spectrum

\begin{eqnarray}
\frac{E^2_{n,l}}{m^2_P}&&= 2n + \sqrt{1  +\left(\, 2l+1\,\right)^2}+\frac{3}{2}\ ,\quad n=0\ ,1\ ,2\ ,3\ ,\dots
\end{eqnarray}

Thus, Hawking radiation at the Planck scale proceeds through single quantum jumps, similar to the decay
of an excited atom. The minimal energy  ground state is

\begin{equation}
E^2_{0,0}=\frac{ 3+2\sqrt{2} }{2}\,m^2_P 
\end{equation}

This is the stable remnant left by the QBH decay process.\\
Finally, it is interesting to notice that, at high angular momentum $l >>1$, QBHs fit on a \emph{linear Regge trajectory}

\begin{equation}
l \simeq \alpha^\prime E^2 +\alpha\left(\,0\,\right)
\end{equation}
with a Regge slope $\alpha^\prime=1/2m_P^2 $ and a negative intercept $\alpha\left(\,0\,\right)=-n-3/4$. This behavior
strongly indicates that the excited states of higher angular momentum are  \emph{stringy} in nature
\cite{Veneziano:2004er,Susskind:1993ws,Horowitz:1996nw,Damour:1999aw} .

\section{Conclusions}
\label{closing}
We introduced  a  general, physically compelling  criterion in order to distinguish between a quantum particle and a QBH. The separations is based on the ratio of the two basic length scales: the Compton wavelength and the  gravitational radius. When their ratio is close to one we are in a genuine quantum gravity regime and the geometric, static Schwarzschild horizon does not provide  a satisfactory description of a QBH. In order to provide a proper description, we introduced a Generalized Uncertainty Principle between the QBH mass and its horizon. Eventually  GUP leads to
an "effective" Scharzschild like-geometry   (\ref{qbh1}) with quantum gravity correction to the Newtonian potential. In spite of this simple looking form, the corresponding metric   ensures that QBH can exist only above the Planck mass. On the thermodynamical side, QBH cannot evaporate completely, but rather ends up as zero temperature Planckian remnant.\\
Then we improved this  semi-classical geometric description  by allowing the horizon to undergo quantum fluctuations. A full quantum formulation, based on the GUP formulated at the beginning of the paper, is  formulated.  The wave equation looks like a Sch\"ordinger-type equation (\ref{weq}) for a particle subjected to an effective potential. The exact solution for the horizon wave equation is obtained and the
mass-energy spectrum is found to be discrete and bounded from below. Interestingly enough  excited states with high angular momentum 
correspond to  linear Regge trajectory displaying  a characteristic string-like behavior .

\end{document}